\documentclass[fontsize=11pt,a4paper]{arxiv-article}
\usepackage{Definitions}
\usepackage{comment}

\usepackage{hyperref}
\graphicspath{{./fig/}}

\preprint{\texttt{YITP-24-175}}

\def\ie{\begin{equation}\begin{aligned}}
\def\fe{\end{aligned}\end{equation}}

\newcommand{\OfficialTitle}{
	Asymptotics\\in the bi-Yang-Baxter Sigma Model
}
\title{\setstretch{1.4}
	{\color{Thoughtless}\textls[-20]{\OfficialTitle}}
}

\hypersetup{pdfauthor={Meer Ashwinkumar,Domenico Orlando, Susanne Reffert, Giacomo Sberveglieri},pdftitle={\OfficialTitle},%
	colorlinks=true,linkcolor=ThoughtYouWere,citecolor=ThoughtYouWere,urlcolor=ThoughtYouWere}

\author{%
	\begin{minipage}{.94\textwidth}
		\begin{center} \dosserif%
			{\small
				\textbf{Meer Ashwinkumar}\textsuperscript{\ding{73}},
				\textbf{Domenico Orlando}\textsuperscript{\ding{72}\ding{73}\ding{71}},
  				\textbf{Susanne Reffert}\textsuperscript{\ding{73}\ding{71}},\\ and 
				\textbf{Giacomo Sberveglieri}\textsuperscript{\ding{73}},
			}
		\end{center}
		\authorBlock{\ding{73}}{\dosserif{}%
			Albert Einstein Center for Fundamental Physics,\\
			Institute for Theoretical Physics, University of Bern,\\
			Sidlerstrasse 5, CH-3012 Bern, Switzerland}
		\authorBlock{\ding{72}}{\dosserif{}%
			INFN sezione di Torino.\\
			via Pietro Giuria 1, 10125 Torino, Italy}
       \authorBlock{\ding{71}}{\dosserif{}%
         Yukawa Institute for Theoretical Physics, Kyoto University,\\
         Kyoto 650-0047, Japan}
	\end{minipage}
}

\date{}

\begin{document}

\numberwithin{equation}{section}

\begin{titlepage}

	%\newgeometry{top=23.1mm,bottom=46.1mm,left=34.6mm,right=34.6mm}

	\maketitle

	\thispagestyle{empty}

	%\vfill
   %\dosserif{}

	\abstract{%
		\normalfont{}\noindent{}%
  Working in a sector of large charge is a powerful tool to analytically access models that are either strongly coupled or otherwise difficult to solve explicitly. In the context of integrable systems, Volin's method is exactly such a large-charge approach. In this note, we apply this method to the bi-Yang--Baxter deformed $SU(2)$ principal chiral model. Our main result is an explicit expression for the free energy density as an asymptotic expansion. We moreover determine the leading non-perturbative effects both via analytic methods and a resurgence analysis. 
    }
	%\vfill

\end{titlepage}

%\restoregeometry{}

\setstretch{1.1}
%%%%%%%%%%%%%%%%%%%%%%%%%%%%%
\tableofcontents

%\listoffigures
\newpage

\section{Introduction}%
\label{sec:intro}

In recent years, the large-charge expansion~\cite{Hellerman:2015nra,Monin:2016jmo,Gaume:2020bmp} has proven to be a powerful tool for accessing models that are otherwise difficult to treat analytically. A prime example are strongly-coupled \acp{cft}.
It allows
one to express the observables as an expansion in inverse powers of a large and fixed global charge.
Also in scenarios where there is already a certain amount of control over the model, working in a large-charge limit brings additional advantages.\footnote{Examples are the large-N limit~\cite{Alvarez-Gaume:2019biu,Dondi:2021buw,Dondi:2022zna}, epsilon expansion~\cite{Badel:2019oxl,Watanabe:2019pdh,Arias-Tamargo:2019xld} or superconformal field theories~\cite{Hellerman:2017sur,Hellerman:2018xpi,Hellerman:2020sqj}.}
One limitation of the technique is encountered in the case of two-dimensional strongly-coupled \acp{cft}.
Because of the special structure of the two-dimensional conformal group, in this case fixing the charge results in the decoupling of a free boson and the rest of the theory remains inaccessible~\cite{Komargodski:2021zzy,Araujo:2021mdm}.

Large-charge methods, however, are not restricted to \acp{cft}.
They can also be used for integrable models, which despite being integrable are often hard to solve explicitly~\cite{Dodelson:2023uuu}.
The reach of well-understood models can be increased by studying their integrable deformations.
The \ac{YB} deformation is an important class of them and has generated a lot of interest~\cite{Klimcik:2002zj,Klimcik:2008eq, Klimcik:2014bta}.

The idea of using a simplifying limit is of course not a novel one and has been widely employed in the past. 
In the integrable context, the \ac{tba}, which applies in the case of large particle number, large system size and fixed density, is exactly such an example. Even so, it is in general hard to solve. 
Volin's method~\cite{Volin:2009tqx,Volin:2009wr} provides an efficient way of
expanding the integral equations.
In technical terms, the idea is to write both the energy and the charge density as function of the size \(2B\) of the support of the density of the rapidity distribution \(\chi(\theta)\).
For large density \(\rho \to \infty\),  \(B\) also diverges, and we can use this fact to express the observables as perturbative expansions in \(1/B\).

\medskip

In this paper, we apply Volin's method to the \ac{bYB} model, which is a two-parameter deformation of the \ac{pcm}. 
Having chosen the charge density as a control parameter, we also need to specify how the 
RG-invariant
parameters of the theory, denoted \(p_1\) and \(p_2\), scale with it (or, equivalently with the parameter \(B\)). 
We consider the limit in which \(p_i \to \infty\), keeping \(\bar p_i = p_i/B\) fixed and large, which corresponds to the limit where the deformation parameters of the theory, denoted $\eta$ and $\zeta$, are small.
Our main result is an explicit derivation of the free energy%
\begin{equation}
    e(\rho) = \frac{\pi}{2} \rho^2 \alpha_{\rho} \varphi(\alpha_{\rho}) =  \frac{\pi}{2} \rho^2 \pqty*{ \alpha_{\rho} +  \alpha_{\rho}^2\left(\frac{1}{2} +\frac{2}{p}\right) + \dots } ,
\end{equation}
where $\varphi(\alpha_{\rho})$ is a convenient adimensional quantity, \(p\) is an appropriate combination of the \(p_i\), and \(\alpha_{\rho}\) is the \(\rho\)-dependent coupling that satisfies
\begin{equation}
  \frac{1}{\alpha_{\rho}} - \frac{1}{2} \log(\alpha_{\rho}) = \log*(4\sqrt{\frac{2\pi}{e}} \left(1-\frac{2}{p}\right)\frac{\rho}{m}) \, ,  
\end{equation}
with \(m\) the mass gap, \emph{i.e.}, the mass of the lightest charged particle.%
\footnote{Here we adhere to the standard thermodynamics conventions and call \emph{free energy} the energy as function of the charge \(Q\) (canonical ensemble). Its Legendre transform is the \emph{grand potential} \(\mathcal{E}(h)\), which is function of the chemical potential \(h\)(grand canonical ensemble).
We use the normalization $\mathcal{E}(h)=E(h)-E(0)$, where in a volume \(V \) and total Euclidean time \(\beta\), the energy $E$ per unit volume is defined as 
\begin{equation*}
    E(h)=-\lim_{V, \beta \to \infty} \frac{1}{V \beta} \log \Tr e^{-\beta(H-hQ)}\,.
\end{equation*}
}
For \(\rho \to \infty\) the coupling \(\alpha_{\rho}\) goes to zero and we can write explicitly the leading terms using the asymptotic expansion of the Lambert \(W\) function, since \(\alpha_p = 2/W( \pi/e ( 8 \rho/m (1 -2/p)^2))\):%\todo{fill the dots}
\begin{equation}
    e(\rho) = \frac{\pi}{2} \rho^2\left(\frac{1}{\log*(\frac{\rho}{m})}+\frac{\log*(\log*(\frac{\rho }{m}))+2-\log(32\pi)+\frac{8}{p}}{2(\log*(\frac{\rho }{m}))^2} +\order*{\log*(\frac{\rho}{m})^{-3}}\right)\,.
\end{equation}

Here, we compute the first non-perturbative contribution as well as its coefficient, and confirm the result via analytic methods. 
After a resurgence analysis, we find for the transseries of $e/\rho^2$ for the \ac{bYB} at large $p$
\begin{equation} 
        \frac{e}{\rho^2} \sim \frac{\pi}{2} \alpha_{\rho} \left(\varphi(\alpha_{\rho})\mp\frac{8i}{e}e^{-\frac{2}{\alpha_{\rho}}} \left(1+\frac{4}{p}\right)+ \order*{e^{-\frac{4}{\alpha_{\rho}}}} +\order*{\frac{1}{p^2}} +\ldots\right)\,,
\end{equation}
where $\varphi(\alpha_{\rho})$ is  computed up to first order in $1/p$.

Both Volin's method and the analytic approach to deducing non-perturbative effects involve the \ac{tba} equations as well as the \ac{wh} decomposition of the \ac{tba} kernel. However, in practice, these methods play complementary roles to one another, allowing us to check either method against the other.

The Wiener-Hopf technique has, in fact, been previously applied to the bi-Yang-Baxter sigma model (in the form of the SS model) in the work of Fateev~\cite{Fateev:1996ea} without taking the aforementioned $p_i\rightarrow \infty $ limit of the RG-invariant parameters, where it was shown that the free energy 
consists only of non-perturbative exponential terms without perturbative series multiplying them.\footnote{To be precise, Fateev first studied the SS model (a theory with three scalar fields with Sine-Gordon-type interactions)  
and showed that particles in this model have an S-matrix involving two Sine-Gordon factors.
The SS model 
and a certain sigma model were shown to have the same free energy in a certain large charge limit. This sigma model was shown to be equivalent to the bi-Yang-Baxter sigma model in \cite{Hoare:2014pna}.} On the other hand, the \ac{bYB} model is a deformation of the bosonic principal chiral model, whose free energy involves non-perturbative exponential terms that are each  multiplied by a perturbative series. It is important to understand how these results are related in the  $p_i\rightarrow \infty $ limit, and this is one of the main motivations of our work. 

We note that there has been previous work by Schepers and Thompson~\cite{Schepers:2020ehn} on resurgence in the bi-Yang-Baxter sigma model.
However, our approach and setup differs significantly from theirs, especially in utilizing Volin's method to generate a long perturbative series for the free energy.
We intend to address the connection between our work and theirs in future studies.

\bigskip
The plan of this note is as follows. In Section~\ref{sec:literature-review}, we briefly review the \ac{bYB} as a \ac{YB} deformation of the \ac{pcm}. We first give its Lagrangian and \acl{rg} equations (Sec.~\ref{sec:LagRG}), and discuss the coupling to a chemical potential, the resulting symmetry structure, and its \ac{tba} equations (Sec.~\ref{sec:TBAchemPot}). Next, we introduce Volin's method (Sec.~\ref{sec:volins-method}).
In Section~\ref{sec:bi-yang-baxter}, we apply the method to the \ac{bYB} model.  We first perform a large $p_i$ expansion of the function $G_+$ that arises from the Wiener-Hopf decomposition of the \ac{tba} kernel (Sec.~\ref{sec:kern-wh-decomp}) and then write the perturbative series for $e/\rho^2$ in a double-scaling limit.
In Section~\ref{sec:TransseriesResurgence}, we use this perturbative series to determine non-perturbative properties of the \ac{bYB} model via a resurgence analysis.
In Section~\ref{sec:perturbative-sigma-model}, we perform a perturbative calculation to verify our earlier results. In Section~\ref{sec:conclusions}, we summarize our findings and present some further directions. In Appendix~\ref{app:first}, we give the precise relationship between the \ac{tba} kernels of the \ac{bYB} model and the \ac{pcm}.

%%%%%%%%%%%%%%%%%%%%%%%%%%%%%%%%%%%%%%%%%%%%%%%
\section{The bi-Yang Baxter model}
\label{sec:literature-review}

\subsection{Lagrangian and RG}\label{sec:LagRG}

Let us start reviewing some of the key aspects of the so-called \ac{bYB} model.
Its Lagrangian reads
\begin{equation}
    \mathcal{L}_{\text{bYB}}=-\frac{1}{a^2_0} \Tr*(g^{-1}\partial_+g\frac{1}{1-\eta \mathcal{R}_g-\zeta \mathcal{R}}g^{-1}\partial_-g )\,.
\end{equation}
The field $g$ takes values in a simple Lie group $G$ with Lie algebra $\text{Lie}(G)=\mathfrak{g}$.
$\mathcal{R}$ is a constant operator on $\mathfrak{g}$, acting in the adjoint representation $\text{Ad}_g X= g X g^{-1}$, and $\mathcal{R}_g=\text{Ad}^{-1}_g \mathcal{R}\text{Ad}_g$.
This model is a deformation of the \ac{pcm}: turning off both deformations, \emph{i.e.} $\eta=\zeta=0$, we find 
\begin{equation}
  \mathcal{L}_{\text{PCM}}=-\frac{1}{a^2_0} \Tr*(g^{-1}\partial_+gg^{-1}\partial_-g)\,.
\end{equation}
The \ac{pcm} has a $G_L\times G_R$ global symmetry.
If we turn on the $\eta$-deformation alone, we break $G_L$ to a subgroup $\widetilde{G}_L$, whose elements commute with the operator $\mathcal{R}$. This is usually called the \ac{YB} deformation of the \ac{pcm}.\footnote{See \emph{e.g.} ref.~\cite{Hoare:2021dix} for a nice review on integrable deformations.}
Turning on both deformations further reduces the global symmetry to $\widetilde{G}_L \times \widetilde{G}_R$, 
where $\widetilde{G}_R$ also commutes with $\mathcal{R}$.

In this work, we will focus on the \ac{bYB} deformation of the $G=SU(2)$ \ac{pcm} with 
\begin{equation}
  \label{eq:R-matrix}
    \mathcal{R}=\begin{pmatrix}
0 & -1 & 0\\
1 & 0 & 0\\
0 & 0 & 0
\end{pmatrix}\,.
\end{equation}
In this setup the original global symmetry $SU(2)_L\times SU(2)_R$ is broken down to $U(1)_L\times U(1)_R$.

The \ac{rg} properties of the \(SU(2)\) \ac{pcm} are well known.
It is an asymptotically free theory and the beta function for the coupling \(a \) is
\begin{equation}
    \beta(a)=\mu \frac{d}{d\mu} a = - \frac{1}{8\pi}a^3 -\frac{1}{64\pi^2}a^5 + \dots\,.
\end{equation}
It is convenient to rewrite it in terms of the coupling $\alpha\equiv a^2/4\pi$:\footnote{In the literature sometimes the label $t$ is used to indicate the \ac{pcm} coupling.
  So in our notation, we would have $\alpha=t/2$ of ref.~\cite{Schepers:2020ehn} or $\alpha=t/4\pi$ of ref.~\cite{Squellari:2014jfa}.}
\begin{equation}
  \beta(\alpha)= - \alpha^2 - \frac{1}{2} \alpha^3 + \dots\, ,
\end{equation}
so that in presence of the deformations the \ac{rg}-flow equations take the simple form~\cite{Sfetsos:2015nya}
\begin{equation}
  \begin{dcases}
    \beta(\alpha) = - \alpha^2 \pqty*{1+\pqty{\eta+\zeta}^2} \pqty*{1+\pqty{\eta-\zeta}^2} + \dots \, , \\
    \beta(\alpha \eta) = 0+\ldots \, ,\\
    \beta(\alpha \zeta) = 0+\ldots \, ,
  \end{dcases}\label{rgfloweq}
\end{equation}
where the ellipsis stands for scheme-dependent terms occurring at order $\alpha^3$ and higher.
In the case of just one coupling constant, only the first two terms are \ac{rg}-independent, while with more than one coupling, only the first term is independent. 
The products $\alpha \eta$ and $\alpha \zeta$ are one-loop \ac{rg} invariants. Our analysis shall utilize the equivalence between the bi-Yang-Baxter sigma model and the SS model studied by Fateev in \cite{Fateev:1996ea}, which holds when $\eta$ and $\zeta$ are imaginary.
For our future convenience we introduce the  quantities
\begin{align}
  \label{eq:p1p2}
  p_1 &= \frac{\pi}{ 2 i\alpha \eta}, & p_2 &= \frac{\pi}{2 i \alpha \zeta}\,,
\end{align}
that are the \ac{rg} invariant parameters of the SS model. The identification between the parameters can be found by comparing \eqref{rgfloweq} with the \ac{rg}-flow equations of the SS model.

\subsection{TBA and coupling with chemical potential}\label{sec:TBAchemPot}

As is well known, the \ac{bYB} model is integrable.
We can couple an external field to a conserved current and write the \acl{tba} equations. %\todo{we need a 1-page review of the \ac{tba} in the appendix, if anything to fix the notation.}
By doing so, as observed by Polyakov and Wiegmann~\cite{Polyakov:1983tt}, we are able to compute the free energy exactly as a function of this external field $h$, which plays the role of chemical potential.
Unfortunately, these equations are too complicated to be solved analytically. However, by taking $h\to \infty$, we can probe the \ac{uv} of the theory and calculate the free energy in perturbation theory as done in~\cite{Polyakov:1983tt,Wiegmann:1985jt,Hasenfratz:1990zz,Hasenfratz:1990ab,Forgacs:1991rs,Forgacs:1991ru,Balog:1992cm,Hollowood:1994np,Evans:1994sy,Evans:1994sv,Evans:1995dn,Fateev:1994dp,Fateev:1994ai,Zarembo:2008hb} and more recently~\cite{Mariño_2019,Marino:2019eym,Abbott:2020mba,Abbott:2020qnl,Marino:2021six,DiPietro:2021yxb,Marino:2021dzn,Bajnok:2021dri,Bajnok:2021zjm,Marino:2022ykm,Bajnok:2022rtu,Bajnok:2022xgx,Schepers:2023dqk,Bajnok:2025mxi}.

For general values of the deformation parameters, our model has a \(U(1)_L \times U(1)_R\) symmetry, so we can couple it to two external fields \(h_L\) and \(h_R\).
The values taken by these fields are fixed by asking for a given charge configuration \((Q_L, Q_R)\) of the corresponding ground state.
Different charge configurations are described by different components of the S-matrix.

In the following, we will use a different representation of the \ac{bYB} deformation in terms of the two-parameter Fateev model (also known as the SS model)~\cite{Fateev:1996ea}, which was shown to be equivalent to the \ac{bYB} model in~\cite{Hoare:2014pna}.
One subtlety in this mapping is related to the fact that Fateev's model is a deformation of the \(O(4)\) $\sigma$-model, and the \(SU(2) \times SU(2)\) symmetry of the undeformed \ac{pcm} model is the double cover of \(SO(4)\):
\begin{equation}
  SO(4) = \frac{SU(2) \times SU(2)}{\{(1,1), (-1,-1)\}}.
\end{equation}
In Fateev's model, the natural (unbroken) charges correspond to the Cartan generators of \(SO(4)\), which are related to \(Q_L\) and \(Q_R\) by
\begin{equation}
  \begin{cases}
    Q_1 = Q_L + Q_R \, , \\
    Q_2 = Q_L - Q_R \,.
  \end{cases}
\end{equation}
The double-cover structure implies that, while \(Q_L\) and \(Q_R \) can be integers or half-integers as representations of \(SU(2) \times SU(2)\), only pairs such that \(Q_1\) and \(Q_2\) are integers are allowed as representations of \(SO(4)\).

In this work we will concentrate on the sector with \(Q_1\) fixed (and large) and \(Q_2 = 0\), which in the undeformed model corresponds to the completely symmetric representation of \(SO(4)\) with \(Q\) boxes, and in terms of \(SU(2) \times SU(2)\), to the ``axial'' \(U(1)\) charge.
Note that this configuration cannot obtained by setting \(h_2 = 0\), simply because the Lagrangian has a non-trivial target-space metric.
In Fateev's words, one needs a ``compensating field''~\cite{Fateev:1996ea}.

Our goal is to compute the energy density \(e \) as function of the particle density \(\rho\) (we are in the canonical ensemble) in the limit in which the number of particles \(Q\) and the length of the system \(L\) go to infinity, at fixed \(\rho = Q/L\).
In the \ac{tba}, this is obtained by computing the density of the rapidity distribution \(\chi(\theta)\) which satisfies the integral equation
\begin{equation} \label{eq:TBAchi}
  \chi(\theta) - \int_{-B}^B \dd{\theta'} K(\theta - \theta' ) \chi(\theta') = m \cosh(\theta) .
\end{equation}
Here, we have indicated explicitly the boundary of the support of \(\chi(\theta)\),
\begin{equation}
   \supp(\chi) \subset (-B,B) \, ,
\end{equation}
and the kernel \(K(\theta)\) is the logarithmic derivative of the S-matrix,
\begin{equation}
  K(\theta) = \frac{1}{2 \pi i} \odv*{\log(S(\theta))}{\theta} .
\end{equation}

The main advantage of the two-parameter model that we have chosen is that its S-matrix factorizes into the product of two copies of the S-matrix for solitons in the sine-Gordon model:

\begin{equation}
  S^{bYB}_{p_1, p_2}(\theta) = - S_{\gamma=p_1}^{S G}(\theta) \otimes S_{\gamma=p_2}^{S G}(\theta) \, ,
  \label{smatt}
\end{equation}
where $p_1$ and $p_2$ were introduced in Eq.~(\ref{eq:p1p2}). As explained by Fateev~\cite{Fateev:1996ea}, this model can be described in terms of two coupled Sine-Gordon models.
When the parameters satisfy $p_1\geq 1$ and $p_2 \geq 1$, there are four different types of particles in the theory. We shall be interested in these ranges for $p_1$ and $p_2$ in what follows.

In our configuration, only particles of type 1 (with \(Q_1 \neq 0 \) and \(Q_2 = 0\)) condense.
This corresponds to pure soliton-soliton scattering \(S_{11}(\theta) = - s_{p_1}(\theta) s_{p_2}(\theta)\), where  
\begin{equation}
  \begin{aligned} 
    s_\gamma^{S G}(\theta) & =\exp*[i \int_{-\infty}^{\infty} \frac{\dd{\omega}}{2 \omega} \frac{\sin (\theta \omega) \sinh (\pi(\gamma-1) \omega / 2)}{\cosh (\pi \omega / 2) \sinh (\pi \gamma \omega / 2)}] \\ & =\exp*[i \int_{-\infty}^{\infty} \frac{\dd{\omega}}{2 \omega} \sin (\theta \omega)(1-\coth(\pi \gamma \omega / 2) \tanh (\pi \omega / 2)) ]\,.
  \end{aligned}
\end{equation}
Thus, the scattering kernel appearing in the \ac{tba} equations will be 
\begin{equation}
  K^{bYB}(\theta)=\frac{1}{2 \pi i} \frac{d}{d \theta} \log*(- s_{p_1}^{S G}(\theta)s_{p_2}^{S G}(\theta))\, ,
\end{equation}
and computing its Fourier transform \(\widetilde{K}^{bYB}(\omega)\), we find
\begin{equation}
  1-\widetilde{K}^{bYB}(\omega)=\frac{1}{2} \tanh (\pi \omega / 2) \frac{\sinh*(\pi\left(p_1+p_2\right) \omega / 2)}{\sinh*(p_1 \pi \omega / 2) \sinh*(p_2 \pi \omega / 2 )}.
\end{equation}

Now that we have the \ac{tba} equations ready, we can prepare to solve them perturbatively at large chemical potential.
To do so, we will use the \acl{wh} method~\cite{hopf1932klasse}.
Its key step is the decomposition of the kernel in terms of the product of two functions, analytic in the upper and lower halves of the complex plane:
\begin{equation}
  1-K(\omega)=\frac{1}{G_{+}(\omega) G_{-}(\omega)}\,.
\end{equation}
For our model, $G^{bYB}_-(\omega)=G^{bYB}_+(-\omega)$ and
\begin{equation} \label{eq:G_bYB}
  G^{bYB}_{+}(\omega)=e^{-i b \omega}\left(\frac{p_1+p_2}{2 p_1 p_2}\right)^{1/2} \frac{\B*(\frac{-i \omega}{2}, \frac{-i \omega}{2})}{\B*(\frac{-i p_1 \omega}{2}, \frac{-i p_2 \omega}{2})}. 
\end{equation}
Here, \(B\) is the beta function $\B(x,y)=\Gamma(x)\Gamma(y)/\Gamma(x+y)$, and \(b\) is the parameter
\ie \label{beta}
b =\frac{1}{2}(p_1 \log p_1+p_2 \log p_2-(p_1+p_2) \log (p_1+p_2) + \log 4)\,,
\fe 
so that $G^{bYB}_{+}(i x) = 1 + \order{1/x}$ for $x \to \infty$.%

%%%%%%%%%%%%%%%%%%%%%%%%%%%%
\subsection{Volin's method}
\label{sec:volins-method}

In this section, we briefly review a powerful technique for extracting long perturbative series starting from the \ac{tba} equations introduced by Volin~\cite{Volin:2009wr,Volin:2009tqx}.\footnote{For an organic introduction, see chapter 6 of ref.~\cite{Reis:2022tni}.} 
The method allows one to
systematically find the ground-state rapidity distributions in the weak-coupling regime corresponding to the limit $B \gg 1$.
We list here the main steps since we will build upon them later.
Volin's method involves computing the resolvent  
\begin{equation}\label{eq:res}
R(\theta)=\int_{-B}^B \dd{\theta^{\prime}} \frac{\chi\left(\theta^{\prime}\right)}{\theta-\theta^{\prime}}\,,
\end{equation}
distinguishing the bulk (\(\theta\) far from \(\pm B\)) and an edge (\(\theta\) close to the branch cuts at \(\pm B\)) limit, both with $B \gg 1$.
Through this approach, we obtain two ansätze.
By ensuring their mutual consistency we find algebraic constraints that lead to a perturbative expansion of the observables.

The bulk limit is the double-scaling limit for which
\begin{equation}
    B,\ \theta \to \infty \qquad \text{with} \quad u\equiv \theta/B \quad \text{finite}.
\end{equation}
In this limit, it is possible to rewrite the \ac{tba} equation~(\ref{eq:TBAchi}) in terms of the resolvent~(\ref{eq:res}) and shift operators.%
\footnote{A shift operator is an operator that acts on a function as $DF(z)=F(z+i)$, so $D\equiv e^{i\partial_z}$.}
From there, an ansatz in terms of a power series in $1/B$ can be deduced for the resolvent.
In models in which $\widetilde{K}(0)=1$ --- which is the case for bosonic systems\footnote{See Section 3.3.6 of~\cite{Reis:2022tni} for a detailed discussion of the distinction between bosonic and fermionic systems.} --- the bulk ansatz takes the form 
\begin{equation}
  \label{bulka}
  R(\theta)=2A \sum_{n, m=0}^{\infty} \sum_{k=0}^{m+n} \frac{c_{n, m, k}(\theta / B)^{k \ \text{mod} \ 2}}{B^{m-n-1/2}\left(\theta^2-B^2\right)^{n+1 / 2}}\left(\log \left(\frac{\theta-B}{\theta+B}\right)\right)^k.
\end{equation} 
We will show that in the limit where $p_i \rightarrow \infty$,  this ansatz is also relevant to the \ac{bYB} model,  even though it is not bosonic (in the sense that $\widetilde{K}(0)=1$) before taking the limit.

In the edge limit, we have
\begin{equation}
    B,\ \theta \to \infty \qquad \text{with} \quad\, z\equiv 2(\theta-B)\quad \text{finite}.
\end{equation}
In this regime it is convenient to work in Laplace space and exploit the \ac{wh} method introduced in the previous section.
The Laplace transform of the resolvent is defined as
\begin{equation}
    \widehat{R}(s)=\int^{i \infty+0}_{-i \infty+0} \frac{\dd{z}}{2\pi i} e^{sz} R(z)\,,
\end{equation}
and we use the ansatz
\begin{equation}
    \widehat{R}(s)= \frac{m e^B G_+(i)}{2} G_+(2is)\left(\frac{1}{s+\frac{1}{2}}+\frac{1}{B s} \sum_{n, m=0}^{\infty} \frac{Q_{n, m}}{B^{m+n} s^n}\right). \label{edgea}
\end{equation}

Now, what is left to do is to match the two ansätze for the resolvent in the overlapping regime of validity of the two double scalings.
By doing so, we will fix the coefficients $c_{n, m, k}$ and $Q_{n, m}$.
The limit for which the matching procedure has to be performed is
\begin{align}\label{eq:matchVolin}
	 B,\ z &\to \infty  && \text{with} & z/B \to 0.
\end{align}
In terms of the dual variable \(s\), this is \(s \to 0\) and we find an expansion in powers of \(s\) and \(\log(s)\) at each order in \(1/B\).
This will be matched, after an inverse Laplace transform, to an expansion of the bulk that in the overlapping limit of Eq.~(\ref{eq:matchVolin}) turns out to be organized as a series in $(z/B)^n\log(z/B)^r$ at each order in $1/B$.
Concretely, following this procedure, we find a sequence of linear equations that fix the coefficients $c_{n, m, k}$ and $Q_{n, m}$ order by order.

From here, using the definitions of the density $\rho$ and the energy density $e$,
\begin{align}
  \rho(B) &= \int_{-B}^B \frac{\dd{\theta}}{2 \pi} \chi(\theta) \, , & e(B) &= m \int_{-B}^B \frac{\dd{\theta}}{2 \pi} \chi(\theta) \cosh(\theta) ,
\end{align}
it is straightforward to obtain
\begin{equation}\label{eetil}
    \rho(B)= m e^B \sqrt{B}\frac{A}{\pi}\left(1+\sum_{m=1}^{\infty} \frac{c_{0,m,0}-c_{0,m,1}}{B^m}\right)=m e^B \sqrt{B}\frac{A}{\pi}\tilde{\rho} \,,
\end{equation}
\begin{equation}
    e(B)=m^2 e^{2B} \frac{A^2}{\pi^2 k^2}\left(1+\sum_{m=0}^{\infty}\frac{1}{B^m}\sum_{t=0}^{m-1}2^{t+1}Q_{t,m-1-s}\right)=m^2 e^{2B} \frac{A^2}{\pi^2 k^2} \tilde{e}\, ,
\end{equation}
showing explicitly that \(B \to \infty\) is a large charge density limit.
The normalization is composed of the constant $k$ which is related to the function $G_+(\omega)$ as
\begin{equation}
	G_+(i \omega)= k/\sqrt{\omega} \left(1+\mathcal{O}(\omega)\right)
\end{equation}
and of the prefactor $A=kG_+(i)\sqrt{2\pi}/4$ -- the same as in the bulk ansatz.
Note that the constant $A$ drops if we consider the ratio $e(\rho)/\rho^2$.

In order to have a series expressed in terms of an actual running coupling, it is best to define a \(\rho\)-dependent coupling $\alpha_{\rho}$ via
\begin{equation} \label{eq:alphaB_RG}
    \frac{1}{\alpha_{\rho}}+ (\xi-1) \log(\alpha_{\rho})= \log \left(\mathfrak{c}\frac{\rho}{m}\right)\,,
\end{equation}
where \(m\) is the mass gap, and \(\xi\) and \(\mathfrak{c}\) are model-dependent constants. The parameter \(\xi\) is related to the \ac{rg} of the model, while \(\mathfrak{c} \propto 1/(kG_+(i))\).

The $\beta$-function of $\alpha_{\rho}$ reads
\begin{equation}
  \beta(\alpha_{\rho}) = -\alpha_{\rho}^2 - \xi \alpha_{\rho}^3 + \dots .
\end{equation}
For \(\rho \to \infty\), \(\alpha_{\rho}\) is infinitesimal and can be used to write weakly-coupled perturbative expansions.
Crucial for the method is the fact that \(1/B\) has a simple power series,
\begin{equation}
  \label{eq:balseries}
  \frac{1}{B} = \sum_{n=1}^{\infty} b_n \alpha_{\rho}^n.
\end{equation}
Finally, a natural adimensional combination to study is $e/\rho^2$.
For bosonic models,  as well as the \ac{bYB} model we study here, one defines
\begin{equation}
  \label{eq:varsigma}
    \alpha_{\rho}\varphi(\alpha_{\rho})\coloneqq k^2\frac{e}{\rho^2}\,,
\end{equation}
so that
\begin{equation}
    \varphi(\alpha_{\rho})=1+\sum_{i=1}c_i\alpha_{\rho}^i\,.
\end{equation}

%%%%%%%%%%%%%%%%%%%%%%%%%%%%%%%%%%%
%\section{Bi-Yang-Baxter and PCM}
\section{Perturbative expansion for small deformations}
\label{sec:bi-yang-baxter}

\subsection{The kernel and the Wiener--Hopf decomposition}
\label{sec:kern-wh-decomp}
The free energy for the \ac{bYB} model was previously derived by Fateev in~\cite{Fateev:1996ea}, where it was shown to consist only of non-perturbative exponential terms without perturbative series multiplying them. As we reviewed above, the \ac{bYB} model is a deformation of the \ac{pcm}, and for the latter model, the free energy can be written in terms of a trans-series that involves non-perturbative exponential terms that are each multiplied by a perturbative series. Our goal is to understand how these results are related in the limit where the RG-invariant parameters are large, \emph{i.e.}, the $p_i\rightarrow \infty$ limit.

It turns out that the expansion of $G^{bYB}_{+}(\omega)$ at large $p_1$ and $p_2$ makes this function amenable to Volin's procedure, allowing us  to write our observables as long perturbative series in $\alpha_{\rho}$ and the deformation parameters.

In this limit, as shown in Appendix~\ref{app:first}, we find
\begin{equation}
  \label{exprex}
   G^{bYB}_{+}(\omega)= G^{PCM}_{+}(\omega)\tfrac{1+\frac{i}{6 (p_1+p_2)\omega}-\frac{2}{ (12(p_1+p_2)\omega)^2}+\mathcal{O}\left(\frac{1}{((p_1+p_2)\omega)^3} \right) + \ldots }{\left(1+\frac{i}{6  p_1 \omega}-\frac{2}{ (12 p_1 \omega)^2}+\mathcal{O}\left(\frac{1}{(p_1 \omega)^3}\right)+\ldots \right)\left(1+\frac{i}{6 p_2\omega}-\frac{2}{ (12 p_2 \omega)^2}+\mathcal{O}\left(\frac{1}{(p_2 \omega)^3}\right)+\ldots\right)},
\end{equation}
where the ellipses denote exponentially suppressed terms,\footnote{The relation~\eqref{exprex} is based on Stirling's approximation, which is an asymptotic series that is itself resurgent. See~\cite{sauzin2021variations}, for example, for further details. } 
and where 
\begin{equation} \label{eq:Gp_PCM}
  G^{PCM}_{+}(\omega)=\frac{e^{-i \omega \log2}}{\sqrt{-\frac{\pi}{2} i\omega}} \frac{\Gamma(1-i\frac{\omega}{2})^2}{\Gamma(1-i\omega)}\,
\end{equation}
is the function $G_+$ for the $SU(2)$ \ac{pcm} (or, equivalently, the $O(4)$ non-linear sigma model).

If we combine the deformation parameters as $x\equiv p_2 / p_1 = \eta/ \zeta$, we can rewrite Eq.~\eqref{exprex} so that  $G^{bYB}_{+}(\omega)$ is expressed in terms of $G^{PCM}_{+}$ together with corrections in powers of $1/p_1$:
\begin{multline}
  \label{gbyb}
  G^{bYB}_{+}(\omega) = G^{PCM}_{+}(\omega) \Bigg( 1-\frac{2 i}{p_1\omega} \pqty*{\frac{1+x(1+x)}{12 x(1+x)}} -\frac{4}{2(p_1\omega)^2} \pqty*{\frac{1+x(1+x)}{12 x(1+x)}}^2   \\ 
  + \order{(p_1\omega)^{-3}}  \Bigg)\,.
\end{multline}
The Stirling approximation used to derive \eqref{exprex} is valid
since we will take $\omega $ to be positive and imaginary, and $p_1$ and $p_2$ to be positive and real.
These conditions on $p_1$ and $p_2$ follow since we defined $\eta$ and $\zeta$ to be imaginary in Section \ref{sec:LagRG}. In fact, for $\zeta=0$, this corresponds to the case of the $\lambda$-deformed principal chiral model
to which the Yang--Baxter sigma model is related via Poisson--Lie T-duality~\cite{Hoare:2015gda,Klimcik:2015gba,Sfetsos:2015nya},
with the deformation parameters related as 
\ie 
\eta=i \frac{1-\lambda}{1+\lambda}.
\fe
A study on the $\lambda$-deformed model using similar techniques to ours has been presented in ref.~\cite{Schepers:2023dqk}.

Different values of the parameter $x$ correspond to different regions of the double deformation.
For example, for $x\to \infty$ and $x=0$, we obtain the one-parameter Yang--Baxter deformation (also called $\eta$-deformation). For $x=1$, we have the configuration with $\eta=\zeta$ --- the so-called critical line for this model that corresponds to the single-parameter $\eta$-deformation of the sigma model on $S^3$ interpreted as the coset $SO(4)/SO(3)$, following the formulation of~\cite{Hoare:2014oua,Delduc:2013qra,}.

\subsection{Perturbative series}

We still need one more step before we can use Volin's technique.
Let us define
\begin{equation}
   p\coloneqq p_1 \pqty*{\frac{12 x(1+x)}{1+x(1+x)}}.
\end{equation}
Now, up to quadratic order in $1/(p_1\omega)$, $G^{bYB}_{+}(\omega)$ in Eq.~\eqref{gbyb} can be expressed as a series in powers of $1/p$, which is small since we are taking $p_1 \gg 1$.\footnote{The expansion of $G^{bYB}_{+}(\omega)$ at cubic and higher orders in $1/p$ will in general have coefficients that have explicit dependence on $x$. }
Since both $p$ and $B$ are taken to be large, we need to specify their relationship.
We choose to consider the double-scaling limit in which
\begin{equation}\label{eq:pbar}
    \bar{p}=\frac{p}{B} \quad \text{is fixed and large.}
\end{equation}
If we rewrite our edge ansatz~\eqref{edgea} in terms of $\bar{p}$, its  analytic structure matches the one of the bulk defined in~\eqref{bulka}  which is the relevant one for bosonic theories with $\widetilde{K}(0)=1$, and we can follow the procedure outlined in Section~\ref{sec:volins-method}.
We compute the resolvent in both the bulk and the edge regime for the \ac{bYB} sigma model, with the coefficients $c_{n,m,k}$ and $Q_{n,m}$ expanded in series of $1/\bar{p}$.
From there, we can compute $\rho$ and $e$ as functions of $1/B$. Recalling the relation between $e$ and $\tilde{e}$ and $\rho$ and $\tilde{\rho}$ defined in~\eqref{eetil}, we obtain, at first order in $1/\bar{p}$,
\begin{align}
  \tilde{e}&=
              1 + \frac{\frac{1}{4} + \frac{1}{3 \, \bar{p}}}{B}+ \frac{\frac{9}{32} + \frac{1}{12 \, \bar{p}}}{B^2} + \frac{\frac{57}{128} + \frac{3}{32 \, \bar{p}}}{B^3}  + \frac{\frac{1875}{2048} + \frac{19}{128 \, \bar{p}} - \frac{27 \, \zeta(3)}{256}}{B^4}+\ldots \,,
  \\
  \tilde{\rho}&
             \!\begin{multlined}[t]
               = 1 + \frac{-\frac{3}{8} + \frac{1}{6 \, \bar{p}}}{B} + \frac{-\frac{15}{128} - \frac{1}{16 \, \bar{p}}}{B^2} + \frac{-\frac{105}{1024} - \frac{5}{256 \, \bar{p}} + \frac{3 \, \zeta(3)}{64}}{B^3} \\
               + \frac{-\frac{4725}{32768} - \frac{35}{2048 \, \bar{p}} + \frac{9 \, \zeta(3)}{128}  + \frac{\zeta(3)}{128 \, \bar{p}}}{B^4}
               +\ldots\,.
             \end{multlined}
\end{align}
In our expansion, we have $k=\sqrt{2/\pi}$. The prefactor $A$ is, instead, proportional to \(G_+(i)\), so it has a non-trivial $p$ dependence. By expanding for large $p$ and keeping corrections up to the first order in $1/p$ we find 
\begin{equation}
 A=\frac{1}{2}\sqrt{\frac{\pi}{2}}\left(1-\frac{2}{p}\right)\,.
\end{equation}

Finally, to compute a series for the energy density in terms of $\alpha_\rho$, we use Eq.~\eqref{eq:alphaB_RG} with $\xi=1/2$ and $\mathfrak{c}=4\sqrt{2\pi/e}(1+2/p)$,\footnote{Also the constant $\mathfrak{c}$ is a function of $p$ through \(G_+(i)\) and we expand it.
}
which allows us to write the parameter \(1/B\) as a series in \(\alpha_{\rho}\):
\begin{equation}
\label{eq:B-as-alpha}
\frac{1}{B}=\sum^{\infty}_{n=1}b_{n}\alpha_{\rho}^n,
\end{equation}
where the first few coefficients are given by
\begin{align}
  b_1 &= 1\,, &
  b_2 &= -\frac{1}{2}\,, &
  b_3 &= \frac{1}{8}+\frac{2}{ \bar{p}}\,,& 
  b_4 &= -\frac{4}{ \bar{p}}\,,& 
  b_5 &= -\frac{19}{384}+\frac{3\zeta(3)}{64}+\frac{4}{\bar{p}}\,.
\end{align}

Putting everything together, we find, up to the first order in $1/\bar{p}$ and using the notation introduced in Eq.~\eqref{eq:varsigma}, the series
\begin{equation}\label{eq:varsigma_pbar}
    \varphi(\alpha_\rho)= 1 + \frac{\alpha_{\rho}}{2} + \alpha_{\rho}^2 \left(\frac{1}{4} + \frac{2}{\bar{p}}\right) + \alpha_{\rho}^3 \left(\frac{5}{16} - \frac{3 \zeta(3)}{32}  \right) + \alpha_{\rho}^4 \left(\frac{53}{96} - \frac{9 \zeta(3)}{64} + \frac{3}{4 \bar{p}} \right)+\ldots\,.
\end{equation}
Or, equivalently, in terms of \(p = B \bar p\), using the expansion in Eq.~\eqref{eq:B-as-alpha},
\begin{multline}\label{eq:varsigma_PT}
    \varphi(\alpha_\rho)= 1 + \alpha_{\rho}\left(\frac{1}{2} +\frac{2}{p}\right)+ \alpha_{\rho}^2 \left(\frac{1}{4} +\frac{1}{p} \right) + \alpha_{\rho}^3 \left(\frac{5}{16} - \frac{3 \zeta(3)}{32}  + \frac{1}{p}\right) \\ + \alpha_{\rho}^4 \left(\frac{53}{96} - \frac{9 \zeta(3)}{64} + \frac{1}{p}\left(2-\frac{3 \zeta(3)}{4} \right) \right)+\ldots\,.
\end{multline}

We have computed this series up to the $39^{\text{th}}$ order in $\alpha_{\rho}$.
As a consistency check, we note that taking $|p| \to \infty$, or in other words, the \ac{rg}-invariant quantity $\alpha_{\rho} \eta \to 0$, our series in Eq.~\eqref{eq:varsigma_PT} becomes the one of the undeformed \ac{pcm}, $\varphi_0(\alpha_\rho)$, as expected.\footnote{It should be noted that this result is not symmetric under $\eta\rightarrow -\eta$ and $\zeta\rightarrow -\zeta$, although the bi-Yang-Baxter sigma model has such a symmetry upon dropping a B-field. This is expected, since the S-matrix \eqref{smatt} does not have such a symmetry. Moreover, the dual SS model studied by Fateev \cite{Fateev:1996ea} also does not have this symmetry even at the classical level, meaning that it is not a symmetry of the quantum theory.  }

%%%%%%%%%%%%%%%%%%%%%%%%%%%%%%%%%%%%%%%%%%
\section{Transseries and resurgence}\label{sec:TransseriesResurgence}

The perturbative series~\eqref{eq:varsigma_PT} is asymptotic in the coupling $\alpha_{\rho}$. Thus, it is just a part of a more generic transseries that also contains non-perturbative terms.
In this section, we will investigate the large-order behavior and the resurgence properties of this series at order $1/p$.

\paragraph{The undeformed model.}

Let us start with the zeroth order% in $1/p$
%the undeformed model
, \emph{i.e.}~the $SU(2)$ \ac{pcm} (or equivalently the $O(4)$ non-linear sigma model). The study of its resurgence structure has been performed in detail in multiple works~\cite{Marino:2019eym,Abbott:2020mba,Abbott:2020qnl,Marino:2021dzn,Bajnok:2021dri,Bajnok:2022xgx}. 
Let us review the main ingredients here. We have 
\begin{equation}
    \frac{e}{\rho^2} \sim \Phi_0(\alpha_\rho,\mathcal{C}^\pm)\,,
\end{equation}
where $\sim$ indicates “asymptotically equivalent to” and $\Phi_0(\alpha_\rho,\mathcal{C}^\pm)$ is a transseries of the form
\begin{equation}
    \Phi_0(\alpha_\rho,\mathcal{C}^\pm)=\frac{\alpha_{\rho}}{k^2}\left(\varphi_0(\alpha_{\rho})+\sum_{i=0}^{\infty}e^{-\frac{2i}{\alpha_{\rho}}}\varphi_0^{(i)}(\alpha_{\rho},\mathcal{C}^\pm)\right)\,.
\end{equation}
It consists of the perturbative series $\varphi_0(\alpha_{\rho})$ and non-perturbative contributions. The $\mathcal{C}^\pm$ are the transseries parameters. In particular, the first correction is entirely due to an \ac{ir} renormalon and is simply
\begin{equation}
    \Phi_0(\alpha_\rho,\mathcal{C}^\pm)=\frac{\alpha_{\rho}}{k^2}\left(\varphi_0(\alpha_{\rho})+\eval*{\mathcal{C}^\pm_{1}}_{\textsc{pcm}}e^{-\frac{2}{\alpha_{\rho}}}+ \order*{e^{-\frac{4}{\alpha_{\rho}}}}\right)\,.
\end{equation}
The $\pm$ is associated to a two-fold ambiguity that compensates the ambiguity present in the resummation prescription. Indeed, if we indicate by $s_\pm$ the lateral Borel resummations we have
\begin{equation}
    \frac{e}{\rho^2} = s_\pm(\Phi_0)(\alpha_\rho;\mathcal{C}^\pm)\,.
\end{equation}
The value of the leading transseries parameter is
\begin{equation} \label{eq:C1_PCM}
    \eval*{\mathcal{C}^\pm_{1}}_{\textsc{pcm}}=\mp \frac{8i}{e}\,.
\end{equation}

In general, the first \ac{ir} renormalon contribution corresponds to a term in the relative ground state energy $\mathcal{E}$ independent of the chemical potential $h$, and thus it is related to the vacuum energy of the model at $h=0$.
This contribution comes from the pole in $\omega=i$ that arises when studying the \ac{tba} equation via the \ac{wh} procedure~\cite{Zamolodchikov:1995xk}. 
It is, therefore, possible to express it entirely in terms of the kernel decomposition function $G_+$~\cite{Marino:2021dzn}. For bosonic models, like the \ac{pcm},
the leading transseries parameter is
\begin{equation}
    \mathcal{C}^\pm_{1}=-\frac{\mathfrak{c}^2k^2}{4\pi}G_+(i)G_-'(i\pm 0)\,.
\end{equation}
In fact, if we plug into this equation the function $G^{PCM}_{+}$ from Eq.~(\ref{eq:Gp_PCM}) and the appropriate values of $\mathfrak{c}$ and $k$ for the $SU(2)$ \ac{pcm} we find the value of Eq.~(\ref{eq:C1_PCM}).
Several numerical checks of this value -- starting from the perturbative series $\varphi_0(\alpha_\rho)$ looking at the large-order behavior of its coefficients or Borel resumming it -- have been performed in the literature~\cite{Abbott:2020mba,Abbott:2020qnl,Marino:2021dzn}.

\paragraph{First-order terms in \(1/p\).}

\begin{figure}
\centering
  \includegraphics[width=0.7\textwidth]{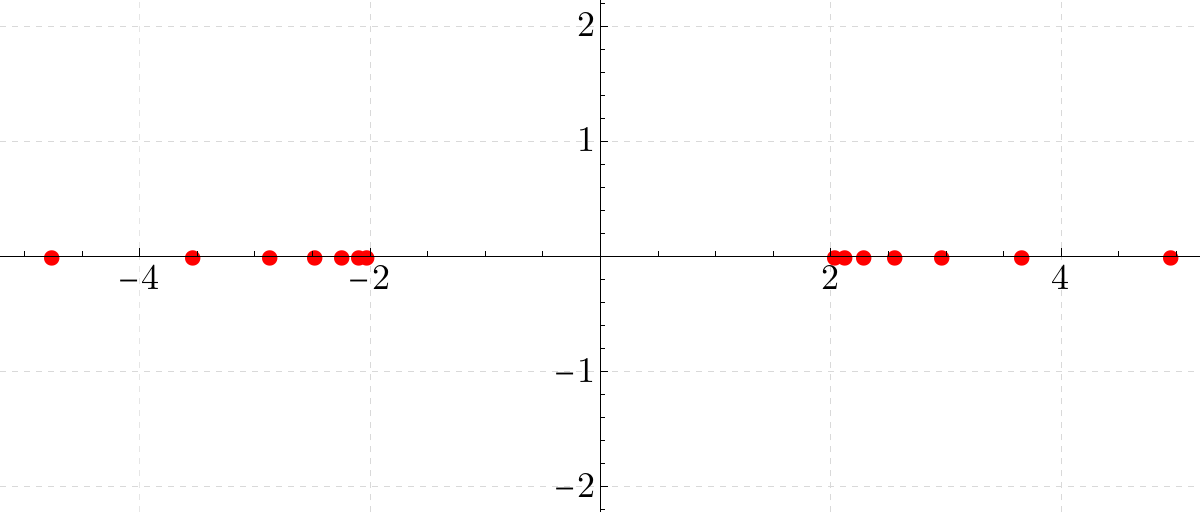}
  \caption{Borel plane with the positions of the singularities for the Padé--Borel [18/18] of our series $\varphi_1(\alpha_\rho)$ in red.}
  \label{fig:LOB_coeffs_phi}
\end{figure}

We can now move on to the first-order corrections in $1/p$ that we have computed. Let us write our perturbative series $\varphi(\alpha_{\rho})$ as 
\begin{equation}
    \varphi(\alpha_{\rho}) = \varphi_0(\alpha_{\rho})+\frac{1}{p}\varphi_1(\alpha_{\rho})\,.
\end{equation}
Unsurprisingly,  the series $\varphi_1(\alpha_{\rho})$ is also divergent in $\alpha_{\rho}$.
Let us study it in more detail.
First of all, we can compute its Padé--Borel transform and plot the poles in the Borel plane, see Figure~\ref{fig:LOB_coeffs_phi}.
As expected, they condense on the real axis, up to $-2$ on the negative side and starting from $2$ on the positive side. 

This indicates the presence of two leading singularities, namely a \ac{uv} and an \ac{ir} renormalon.
With this in mind, we can study the large-order behavior of the coefficients of $\varphi_1(\alpha_{\rho})=\sum_{i}h_i\alpha_\rho^i$. It is given by 
\begin{equation}
    h_n \approx 2^{-n} \tilde{S}_+ \Gamma(n+b_+) \left(1+\order*{n^{-1}}\right)+(-2)^{-n} \tilde{S}_- \Gamma(n+b_-) \left(1+\order*{n^{-1}}\right)\,.
\end{equation}
The parameters $b_+$ and $\tilde{S}_+$ are related to the discontinuity across the positive real line and,  in turn, to the first correction in the transseries.
We have
\begin{equation}
    \disc s(\varphi_1) (\alpha_{\rho}) \approx 2\pi i \tilde{S}_+ 2^{b_+} \alpha_\rho^{-b_+}  e^{-\frac{2}{\alpha_\rho}} + \dots\,.
\end{equation}
Thus, the Stokes parameter $\tilde{\mathsf{S}}_1=-i(\tilde{\mathcal{C}}^{-}_1-\tilde{\mathcal{C}}^{+}_1)$ is
\begin{equation}
    \tilde{\mathsf{S}}_1 = 2\pi \tilde{S}_+ 2^{b_+},
\end{equation}
and, finally,
\begin{equation} \label{eq:SandC}
    \tilde{S}_+ = \frac{2^{-b_+}}{2\pi i}(\tilde{\mathcal{C}}^{-}_1-\tilde{\mathcal{C}}^{+}_1)\,.
\end{equation}
%%%%%%%%%%%%%%%

\begin{figure}
\centering
  \includegraphics[width=0.77\textwidth]{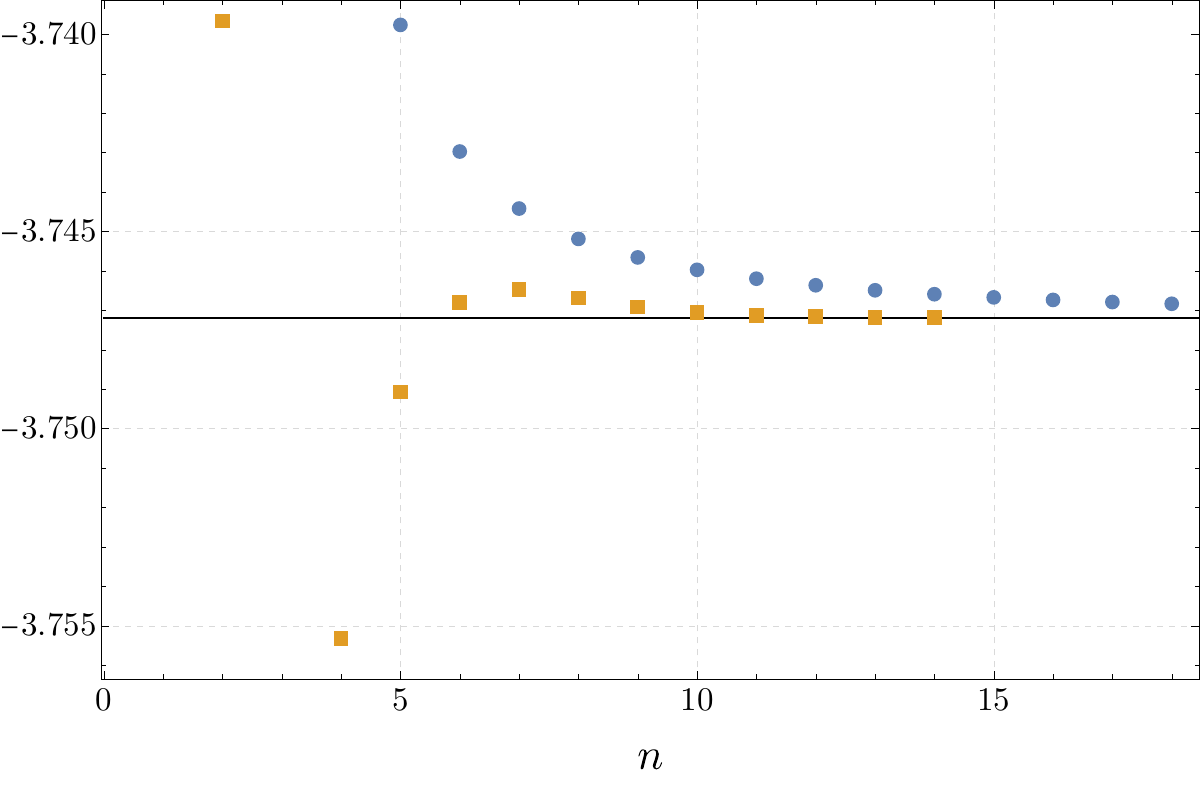}
  \caption{The sequences $s_n$ and its $4^{th}$ Richardson transform $s(4)_n$ in blue and orange, respectively. They approach the solid black line at the value $S_+=32/(e\pi)$.}
  \label{fig:LOB_coeffs_sn}
\end{figure}
%%%%%%%%%%%%%%%%%%%%%%

Having computed the first 39 coefficients $h_m$, we can extract a numerical estimate for them.
Defining
\begin{equation}
    k_n\coloneqq\frac{2^n h_{2n+1}}{(2n+1)!}+\frac{2^{n-1} h_{2n}}{(2n)!}\,,
\end{equation}
we find that
\begin{equation}
    b_n=n \log\left(\frac{k_{n+1}}{k_n}\right)+1
\end{equation}
will approach $b_+$ for $n \gg 1$.
Once a good estimate for it is obtained, one can compute an estimate of $\tilde{S}_+$ via the combination
\begin{equation}
    s_n\coloneqq\frac{2^n h_{2n+1}}{\Gamma(2n+1+b_+)}+\frac{2^{n-1} h_{2n}}{\Gamma(2n+b_+)}\,
\end{equation}
for $n \gg 1$. Moreover, we can use the Richardson transform to accelerate the convergence of these sequences. 
Starting from our coefficients, we find for the $b$-sequence and for its fourth Richardson transform
\begin{align}
    b_{17} & =0.0436\dots\,,\\
    b^{(4)}_{13} & =0.0000221\ldots\,.
    %b^{(6)}_{11} & =9.17\ldots \cdot 10^{-6}\,.
\end{align}
They point towards a value of $b_+=0$. Computing the sequence $s_n$ with $b_+=0$ we find
\begin{equation}
    s_{18} =3.7468\dots\,.
\end{equation}

%3.747189218
%What is this value?? connected to C... Insert figures...
Since we have the analytic form of the transseries parameter $\mathcal{C}^\pm_{1}$ we can use it to compute the correction at the order $1/p$ we are interested in. Plugging in the values for the \ac{bYB} model expanded at large $p$, we find at first order in $1/p$
\begin{equation} 
    \eval*{\mathcal{C}^\pm_{1}}_{\ac{bYB}}=\eval*{\mathcal{C}^\pm_{1}}_{\ac{pcm}}\left(1+\frac{4}{p}\right)=\eval*{\mathcal{C}^\pm_{1}}_{\ac{pcm}}+\frac{1}{p}\tilde{\mathcal{C}}^\pm_{1}=\mp \frac{8i}{e}\left(1+\frac{4}{p}\right)\,.
\end{equation}
In particular, the contributions from $G_+(i)$ and $G_-'(i\pm 0)$ cancel, and we are left with the one coming from $\mathfrak{c}^2$. We can now compare this number with the one coming from the study of the large-order behavior of $\varphi_1(\alpha_{\rho})$. 
Using Eq.~(\ref{eq:SandC}) with $b_+=0$, we find 
\begin{equation}
    \tilde{S}_+=\frac{32}{e\pi}\,.
\end{equation}
Computing the difference with the last values of our sequence $s_n$ and its Richardson transform, we have
\begin{align}
    s_{18}-\tilde{S}_+ & =0.0436\dots\,,\\
    s^{(4)}_{14}-\tilde{S}_+ & =-5.65\ldots \cdot 10^{-6}\,.
\end{align}
This indicates good agreement between the prediction from the perturbative series we have calculated and the analytical value, see Figure~\ref{fig:LOB_coeffs_sn}.

Let us collect our findings for the transseries of $e/\rho^2$ for the \ac{bYB} at large $p$:
\begin{equation} \label{eq:transseries_end}
        \frac{e}{\rho^2} \sim \frac{\pi\alpha_{\rho}}{2}\left(\varphi(\alpha_{\rho}) \mp \frac{8i}{e} e^{-\frac{2}{\alpha_{\rho}}} \left(1+\frac{4}{p}\right)+ \order*{e^{-\frac{4}{\alpha_{\rho}}}} +\order*{\frac{1}{p^2}} +\ldots\right)\,,
\end{equation}
where $\varphi(\alpha_{\rho})$ is the series~\eqref{eq:varsigma_PT}, computed up to first order in $1/p$. 
Calculating further orders in this parameter can be achieved by expanding the kernel decomposition function $G_+$ up to higher orders, keeping in mind that in general, there would also be dependence on $x\equiv p_2/p_1$ at cubic and higher orders in $1/p$. This could be interesting in order to further study the nature of the series in $1/p$ and to investigate the corrections exponentially suppressed in $p$, whose presence is indicated by the ellipsis in Eq.~(\ref{eq:transseries_end}).
We will discuss them in the remaining part of this section.

One of the main goals of this work is to understand the relationship between the free
energy in the \ac{pcm}, which includes a perturbative series, and that of the \ac{bYB}
model, which does not, by exploring the limit from the latter to the former. We have found that in the limit where the deformation parameter $1/p$ (which is $1/p_1$ in the case of a single deformation) is small, the perturbative series indeed approaches that of the \ac{pcm}. We only derived perturbative corrections in $1/p$ up to linear order, but further perturbative corrections at higher orders are expected. A resurgence analysis for a perturbative series in the deformation parameter is expected to give rise to the non-perturbative corrections of the type $e^{-\pi \abs{p_1}}$ and $e^{-\pi \abs{p_2}}$. We leave a detailed analysis for future work.

%%%%%%%%%%%%%%%%%%%%%%%%%%%%%%%%%%%%%%%%%%%%%%%
\section{Perturbative expansion from the sigma model }
\label{sec:perturbative-sigma-model}

In this section we will use perturbative methods to confirm the prediction that there are no $\eta$-dependent terms in the free energy of the \ac{YB} sigma model
up to one loop when working at leading order in $\eta$, as we observed in Eq.~\eqref{eq:varsigma_PT} (recall the relation between $\eta$ and $p$ given in \eqref{eq:p1p2}).

We will study the \ac{YB} sigma model in the form 
\begin{equation}
      \mathcal{L}_{\text{YB}}=-\frac{1}{4 \pi \alpha} \Tr*(g^{-1}\partial_+g\frac{1}{1 - \eta \mathcal{R}_g}g^{-1}\partial_-g )=-\frac{1}{4 \pi \alpha} \Tr*(\partial_+gg^{-1}\frac{1}{1 - \eta \mathcal{R}}\partial_-g g^{-1})\,
\end{equation}
for $g\in SU(2)$ and \(\mathcal{R}\) given in Eq.~(\ref{eq:R-matrix}).
We want to study the system in the weak-coupling regime, so we need to write the action of the operator \(\mathcal{R}\) as explicitly as possible.
Let \(T_a = i \sigma_a\), \(a = 1, 2, 3\) be the generators of the algebra \(su(2)\).
First we decompose
\begin{equation}
   \frac{1}{1-\eta \mathcal{R}}=\gamma \mathcal{A}_{\gamma}+\sqrt{\gamma(1-\gamma)} \mathcal{R},    
\end{equation}
where \(\gamma = 1/(1+\eta^{2})\), and the operators \(\mathcal{A}\) and \(\mathcal{R}\) act on the \(T_{1}\) and \(T_2\) as a rescaling and a rotation:
\begin{align}
  \mathcal{A} &: (T_1, T_2, T_3) \mapsto \pqty*{\frac{1}{\gamma} T_1, \frac{1}{\gamma} T_2, T_3}, \\
  \mathcal{R} &: (T_1, T_2, T_3) \mapsto ( T_2, - T_1, 0) .
\end{align}
We would like to express the action in sigma-model form up to linear order in $\eta$. To this end we employ the parametrization 
\begin{equation}
  g = \exp*(\frac{\varphi+\phi}{2} T_3) \exp*(\theta T_1) \exp*(\frac{\varphi-\phi}{2} T_3 ).
\end{equation}

The $\mathcal{A}$-dependent part of the action can be expressed in sigma-model form with the target-space metric of a squashed sphere:
\begin{equation}
  G = \frac{1}{2 \pi \alpha} \pqty*{ \dd{\theta}^2 + \cos^2(\theta)\dd{\varphi}^2 + \sin^2(\theta) \dd{\phi}^2 + \pqty*{\gamma - 1 } \pqty*{\cos^2(\theta) \dd{\varphi} - \sin^2(\theta) \dd{\phi}}^2}\,.
\end{equation}
Since the dependence on the deformation is linear in $\gamma$, it is also quadratic in $\eta$, and at leading order in \(\eta\) the contribution from the metric to the free energy is the same as in the undeformed \ac{pcm}. 

The $\mathcal{R}$-dependent part of the action leads to a \(B\)-field term of the form 
\begin{equation}
  \int B = -\frac{\eta}{2 \pi \alpha} \int \dd^2{x} \epsilon^{\mu \nu}\partial_{\mu}\theta \partial_{\nu}(\varphi-\phi)\cos(\theta) \sin(\theta ),
\end{equation}
where the Levi--Civita tensor has the components $\epsilon^{+-} = -\epsilon^{-+} = 1$. This $B$-field term is a total derivative, and vanishes unless there is nontrivial holonomy of fields along a cycle of the torus worldsheet. 
Generically, this term will give corrections of order \(\eta\), and therefore we expect that it contributes nontrivially due to the presence of terms odd in $\eta$ in \eqref{eq:varsigma_PT}. 
Here we want to zoom in on the leading-\(\alpha\) contribution, which is obtained by expanding in powers of the field $\theta$:
\begin{equation}
  \int B = -\frac{\eta}{2 \pi \alpha} \int \dd^2{x} \epsilon^{\mu \nu} \theta \partial_{\mu}\theta \partial_{\nu}(\varphi-\phi).
\end{equation}

Now we need to couple our system to a chemical potential so that the ground state has the same charge configuration as described in Sec.~\ref{sec:TBAchemPot}, which in terms of \(SU(2) \times SU(2)\) generators is  \((Q/2, Q/2)\).
At the order at which we are working this is obtained by simply replacing the derivative in the time direction with
\ie 
\partial_0 g \rightarrow \partial_0 g + h(Qg +gQ),
\fe 
where $Q$ has the form
\begin{equation}
  Q= \frac{1}{2} T_3 = \begin{pmatrix}
    \frac{i}{2} & 0 \\
    0 & -\frac{i}{2}
  \end{pmatrix}.
\end{equation}
In our parametrization, only the field \(\varphi\) transforms (non-linearly) under this \(U(1)\), so the modification to the action is just a shift in the time derivative of $\varphi$ by $h$.
The contribution from the B-field involving the chemical potential is thus 
\begin{equation}
 \int B =  \frac{1}{2 \pi \alpha} h \eta \int \dd^2{x} \theta \partial_{1}\theta  .
\end{equation}
To quadratic order in the fields and leading order in $\eta$, the Lagrangian is then 
\ie 
\Lag = \frac{1}{2 \pi \alpha} \pqty*{\partial_+\theta \partial_- \theta + \partial_+\phi \partial_-\phi +\partial_+\varphi \partial_-\varphi  - h^2\theta^2 +h^2 + \eta h \theta \partial_1\theta}.
\fe 

We can now compute the one-loop contribution to the grand potential by computing the determinant of the kinetic operator
\begin{equation}
   M= \frac{1}{\mu^2}\begin{pmatrix}
-\partial^2 - h^2 +\eta h \partial_1 & 0  & 0 \\
0 & -\partial^2 & 0  \\ 0 & 0 & -\partial^2 
\end{pmatrix} \, ,
\end{equation}
for example, by way of zeta function regularization 
\begin{equation}
  \log \operatorname{det} M=-\zeta_M^{\prime}(0)  \, .
\end{equation}
Here, we have introduced the mass scale $\mu$ (to make the eigenvalues of $M$ dimensionless) and the zeta function is
\begin{equation}
  \zeta_M(s) = \Tr(M^{-s}) =  \frac{1}{\Gamma(s)} \int_0^{\infty} \frac{\dd{t}}{t}t^s \Tr*(e^{-t M}) \, .  
\end{equation}

We only need to consider the $\theta$ field, as in our parametrization the $\varphi$ and $\phi$ fields each have the kinetic operator of a free massless boson which contributes a constant.
The one-loop contribution coming from the kinetic operator of $\theta$ then arises from the zeta function
\begin{equation}
  \zeta_M(s)=\frac{ V \mu^{2 s}}{\Gamma(s)} \int_0^{\infty} \frac{\dd{t}}{t} t^{s} \int \frac{\dd^2{k}}{(2 \pi)^2} e^{-t\left(k^2 + i\eta h k_1 -h^2\right)} ,
\end{equation}
where $V$ is the regularized volume of two-dimensional spacetime, and $k$ are spacetime momenta.
Now, the $\eta$-dependent kinetic term can be written as 
\begin{equation}
  -\del_0^2{} -\del_1^2{} + \eta h \del_1{} - h^2 = - \partial_0^2{} - \pqty*{\partial_1{}- \frac{ h\eta}{2}}^2 +\frac{\eta^2h^2}{4} -h^2,
\end{equation}
and the Fourier transform of this quantity is $k_0^2 +(k_1 + ih\eta /2 )^2 + \eta^2h^2 / 4 -h^2$ .

The \(k_1\) integral is over the whole real line, so it is invariant under the shift \(k_1 \to k_1 - i h \eta/2\).
It follows that the \(\zeta\) function is
\begin{equation}
  \zeta_M(s)=\frac{ V \mu^{2 s}}{\Gamma(s)} \int_0^{\infty} \frac{\dd{t}}{t} t^{s} e^{h^2 t} \int \frac{\dd^2{k}}{(2 \pi)^2} e^{-t k^2} + \order{\eta^2} = \frac{V \mu^{2s}(ih)^{2-2s}}{4\pi (s-1)} + \order{\eta^2} .
\end{equation}
This expression has manifestly no contribution of order \(\order{\eta}\).
In other words, at this order in \(\alpha\), the one-loop partition function (and hence the free energy) is the same as for the undeformed \ac{pcm}.
This result is in agreement with the prediction of \(\eta\)-independence of the free energy obtained using Volin's technique, given in Eq.~\eqref{eq:varsigma_PT}, in which the leading deformation scales as \(\order{\alpha^3 \eta}\) (recall that the free energy is proportional to $\alpha \varphi (\alpha)$).

%%%%%%%%%%%%%%%%%%%%%%%%%%%
\section{Conclusions}
\label{sec:conclusions}

Volin's method is an example of a large-charge expansion. It can be applied in the limit of large particle number and large system size and fixed and large density.

In this note, we have used it to compute the free energy density for the integrable \acl{bYB} model in the limit of small deformation parameters as a function of the charge density in terms of an expansion in a charge-dependent coupling which goes to zero for large charge density.
The \ac{bYB} model, being a \ac{YB} deformation of the $SU(2)$ \ac{pcm}, for which similar results have already been found in the literature~\cite{Marino:2019eym,Abbott:2020mba,Abbott:2020qnl,Marino:2021dzn,Bajnok:2021dri,Bajnok:2022xgx}, is a representative of an important class of integrable deformations which have met with a high interest in the community over the last ten years.

In order to be able to successfully apply Volin's method, we had to first  perform a \acl{wh} decomposition, and work in a \emph{double-scaling} limit, where the \(B\) parameter measuring the charge density is taken to be large with the ratio \(\bar p = p/B\) held fixed and large.
We were able to verify our predictions via an independent perturbative computation. We furthermore performed the resurgence analysis to find the non-perturbative behavior of the free energy, going beyond our original perturbative result. Notably, we computed the first non-perturbative correction and its coefficient.

\medskip
Based on this work, there is a number of open questions that we would like to address.
\begin{itemize}
\item One could further explore the relationship between our integrability-based result and the perturbative analysis of the model which we have outlined here.
\item We have studied the simplest possible charge configuration, corresponding to fixing the completely symmetric representation of \(O(4)\). However different charge configurations are possible, describing other sectors of the theory which are in principle accessible with similar techniques.
\item It would be interesting to include higher-order corrections in the deformations $p_1$ and \(p_2\), ideally leading to a double asymptotic expansion in \(\alpha\) and \(p\).
\item One could perform a more systematic study of the non-perturbative physics in terms of the analytic structure of the function \(G_+\) in the \ac{wh} decomposition, as done for example in~\cite{Marino:2021dzn,Bajnok:2022xgx,Bajnok:2025mxi}.
\item One could go beyond the $SU(2)$ \ac{pcm} and try to generalize the construction to the $SU(N)$ case, which have a much richer zoology of sectors and deformations.
\item It would be interesting to understand the relation between our results and those obtained by Schepers and Thompson~\cite{Schepers:2020ehn}.  
\end{itemize}

We leave these points for future investigation.

\section*{Acknowledgments}

\begin{small} \dosserif
  We would like to thank T.~Reis and D.~Thompson for illuminating discussions, as well as N.~Dondi and M.~Serone for detailed feedback on a previous version of the manuscript.
  The work of S.R. and G.S. was supported by the Swiss National Science Foundation under grant number 200021\_219267.
  D.O. and S.R. gratefully acknowledge support from the Yukawa Institute for Theoretical Physics at Kyoto University, as well as from the Simons Center for Geometry and Physics at Stony Brook University, where some of the research for this paper was performed.
\end{small}

%%%%%%%%%%%%%%%%%%%%%%%%%%%%%%%%%%%%%%%%%%%%%%%%%%%%%%
\appendix

\section{Relating $G^{bYB}_+$ and $G^{PCM}_+$}%
\label{app:first}

In this appendix
we shall relate $G^{bYB}_+$ and $G^{PCM}_+$ that appear in the \acl{wh} decomposition of the \ac{tba} kernels for the \ac{bYB} sigma model and \ac{pcm}.
For the sake of convenience, we shall work with the quantities $G^{bYB}_+(2is)$ and $G^{PCM}_+(2is)$, defined respectively as 
\begin{equation}
  \label{a.1}
  G^{bYB}_{+}(2is) = e^{2 b s}\left(\frac{p_1+p_2}{2 p_1 p_2}\right)^{\frac{1}{2}} \frac{\B(s, s)}{\B*(p_1 s, p_2 s)},
\end{equation}
where $b$ is defined in~\eqref{beta} and  $\B(x,y)=(\Gamma(x)\Gamma(y))/\Gamma(x+y)$, and
\begin{equation}
  \label{a.2}
  G^{PCM}_{+}(2is) = \frac{1}{2 \sqrt{\frac{\pi}{4}}} e^{-2 s \log \frac{1}{2}} \frac{\Gamma(s+1) \Gamma(s+1)}{\Gamma(2 s+1)} \frac{1}{\sqrt{s}}.
\end{equation}

To achieve this, we recall Stirling's approximation for the gamma function,
\ie 
\Gamma(z) \sim \sqrt{2 \pi} z^{z-\frac{1}{2}} e^{-z}\left(1+\frac{1}{12 z}+\frac{1}{288 z^2}-\frac{139}{51840 z^3}-\frac{571}{2488320 z^4}+O\left(\frac{1}{z^5}\right)\right).
\fe
The series multiplying $\sqrt{2 \pi} z^{z-\frac{1}{2}} e^{-z}$ is an asymptotic series, that can be completed into a transseries. Based on this expansion, let us define
\ie 
\frac{\Gamma(z)}{\sqrt{2 \pi} z^{z-\frac{1}{2}} e^{-z}} =\lambda(z),
\fe 
where $\lambda(z)$ is a transseries in $z$.
Using the definition of the beta function, one finds that
\begin{equation}
  \frac{1}{\B(p_1 s, p_2 s) }=  \frac{1}{\sqrt{2\pi }} \frac{(p_1+ p_2 )^{p_1s + p_2 s -\frac{1}{2}}}{(p_1)^{p_1s  -\frac{1}{2}}(p_2)^{p_2s  -\frac{1}{2}}} \sqrt{s} \left( \frac{\lambda (p_1s + p_2 s) }{\lambda(p_1 s ) \lambda(p_2 s )} \right).
\end{equation}
It then follows that~\eqref{a.1} can be expressed as 
\ie \label{bbbyb}
G^{bYB}_{+}(2is) = e^{2 b s} \frac{1}{\sqrt{2}} \frac{\Gamma(s) \Gamma(s) }{\Gamma(2s)}\frac{1}{\sqrt{2\pi}}\frac{(p_1+ p_2 )^{p_1s + p_2 s }}{(p_1)^{p_1s  }(p_2)^{p_2s  }} \sqrt{s}\left( \frac{\lambda (p_1s + p_2 s) }{\lambda(p_1 s ) \lambda(p_2 s )} \right).
\fe
Using the expression~\eqref{beta} for $b$, we have
\ie e^{2b s}= \frac{p_1^{p_1 s} p_2^{p_2 s}}{\left(p_1+p_2\right)^{(p_1+p_2)s} } e^{-2\left(\log \frac{1}{2}\right) s}.\fe 
Substituting this expression into~\eqref{bbbyb}, we obtain 
\ie \label{bbbyb2}
G^{bYB}_{+}(2is) =\frac{1}{\sqrt{2}} \frac{\Gamma(s) \Gamma(s) }{\Gamma(2s)}  \frac{1}{\sqrt{2\pi}} \sqrt{s} e^{-2\left(\log \frac{1}{2}\right) s}\left( \frac{\lambda (p_1s + p_2 s) }{\lambda(p_1 s ) \lambda(p_2 s )} \right).
\fe
Finally, using the identities
\ie
\Gamma(s+1) &=s \Gamma(s),\\
\Gamma(2s+1) &=2s \Gamma(2s),
\fe
we find 
\ie \label{bbbyb3}
G^{bYB}_{+}(2is) =\frac{1}{\sqrt{2}} (2) \frac{\Gamma(s+1) \Gamma(s+1) }{\Gamma(2s+1)}  \frac{1}{\sqrt{2\pi}} \frac{1}{\sqrt{s}}e^{-2\left(\log \frac{1}{2}\right) s}\left( \frac{\lambda (p_1s + p_2 s) }{\lambda(p_1 s ) \lambda(p_2 s )} \right),
\fe
and, using~\eqref{a.2},
\ie 
G^{bYB}_{+}(2is) = G^{PCM}_{+}(2is) \left( \frac{\lambda (p_1s + p_2 s) }{\lambda(p_1 s ) \lambda(p_2 s )} \right) .
\fe
% an be equivalently stated as 
% \ie \label{bbbyb4}
% G_+^{Bi-YB}(s)=G_+^{PCM}(s)\left( \frac{\lambda \big(\frac{p_1s}{2i} + \frac{p_2 s}{2i}\big) }{\lambda\big(\frac{p_1 s}{2i} \big) \lambda\big(\frac{p_2 s}{2i} \big)} \right).
% \fe
This is equivalent to the result quoted in~\eqref{exprex}.
% \newpage
\setstretch{1}
\printbibliography{}

\end{document}